\documentclass[sigplan,10pt]{acmart}
\copyrightyear{2024}
\acmYear{2024}
\setcopyright{cc}
\acmConference[PaPoC '24]{The 11th Workshop on Principles and Practice of Consistency for Distributed Data}{April 22, 2024}{Athens, Greece}
\acmBooktitle{The 11th Workshop on Principles and Practice of Consistency for Distributed Data (PaPoC '24), April 22, 2024, Athens, Greece}
\acmDOI{10.1145/3642976.3653030}
\acmISBN{979-8-4007-0544-1/24/04}

\usepackage{listings}
\usepackage{tikz}
\usetikzlibrary{arrows.meta}
\usetikzlibrary{positioning, fit}
\usepackage{svg}
\usepackage{xcolor}
\usepackage{colortbl}
\usepackage{algorithm}
\usepackage{multicol}
\usepackage{breakcites}
\usepackage{soul}
\usepackage[noend]{algpseudocode}

\usetikzlibrary{decorations.pathreplacing,calc}

\makeatletter
\newcommand*{\algrule}[1][\algorithmicindent]{%
  \makebox[#1][l]{%
    \hspace*{.2em}% <------------- This is where the rule starts from
    \vrule height .75\baselineskip depth .25\baselineskip
  }
}

\newcount\ALG@printindent@tempcnta
\def\ALG@printindent{%
    \ifnum \theALG@nested>0% is there anything to print
    \ifx\ALG@text\ALG@x@notext% is this an end group without any text?
    \else
    \unskip
    \ALG@printindent@tempcnta=1
    \loop
    \algrule[\csname ALG@ind@\the\ALG@printindent@tempcnta\endcsname]%
    \advance \ALG@printindent@tempcnta 1
    \ifnum \ALG@printindent@tempcnta<\numexpr\theALG@nested+1\relax
    \repeat
    \fi
    \fi
}
\patchcmd{\ALG@doentity}{\noindent\hskip\ALG@tlm}{\ALG@printindent}{}{\errmessage{failed to patch}}
\patchcmd{\ALG@doentity}{\item[]\nointerlineskip}{}{}{} % no spurious vertical space
\makeatother

\definecolor{insertcolor}{rgb}{0, 0.5, 0}
\definecolor{deletecolor}{rgb}{1, 0, 0}
\definecolor{movecolor}{rgb}{0.5, 0, 1}
\definecolor{networkcolor}{rgb}{0, 0.5, 1}

\definecolor{opset1}{rgb}{0, 0.5, 1}
\definecolor{opset2}{rgb}{0.5, 0, 1}
\definecolor{opset3}{rgb}{0, 0.5, 0}
\definecolor{opset4}{rgb}{1, 0, 0}

\begin{document}
%-------------------------------------------------------------------------------

%-------------------------------------------------------------------------------
\begin{abstract}
%-------------------------------------------------------------------------------

Conflict-Free Replicated Data Types (CRDTs) for JSON allow users to concurrently update a JSON document and automatically merge the updates into a consistent state. Moving a subtree in a map or reordering elements in a list within a JSON CRDT is challenging: naive merge algorithms may introduce unexpected results such as duplicates or cycles. In this paper, we introduce an algorithm for move operations in a JSON CRDT that handles the interaction with concurrent non-move operations, and uses novel optimisations to improve performance. We plan to integrate this algorithm into the Automerge CRDT library. 

\end{abstract}

\title{Extending JSON CRDTs with Move Operations}
\author{Liangrun Da}
\email{liangrun.da@tum.de}
\affiliation{%
  \institution{Technical University of Munich}
  \city{Munich}
  \country{Germany}
}

\author{Martin Kleppmann}
\email{martin@kleppmann.com}
\affiliation{%
  \institution{University of Cambridge}
  \city{Cambridge}
  \country{UK}
}




\maketitle

%-------------------------------------------------------------------------------
\section{Introduction}
%-------------------------------------------------------------------------------

\begin{sloppypar}Automerge~\cite{automerge} is an implementation of Conflict-Free Replicated Data Type (CRDT)~\cite{shapiro2011conflict}, which allows concurrent changes to data on different devices to be merged automatically without requiring any central server. It is used in the development of local-first software~\cite{kleppmann2019local}, which includes applications such as collaborative drawing, text editing, and more.\end{sloppypar}

Automerge uses a \emph{document} as its data model, which can be viewed as a JSON data type. A document can be accessed locally through operations such as \emph{get}, \emph{put}, and \emph{delete}, and any modifications are replicated to other devices. Many applications require reordering elements in a list, e.g. when using drag-and-drop to change the order of a to-do list. Moreover, some applications need to move a subtree to a new parent node: for example, in a document representing a file system tree, moving a file or directory from one location to another is a very common operation.

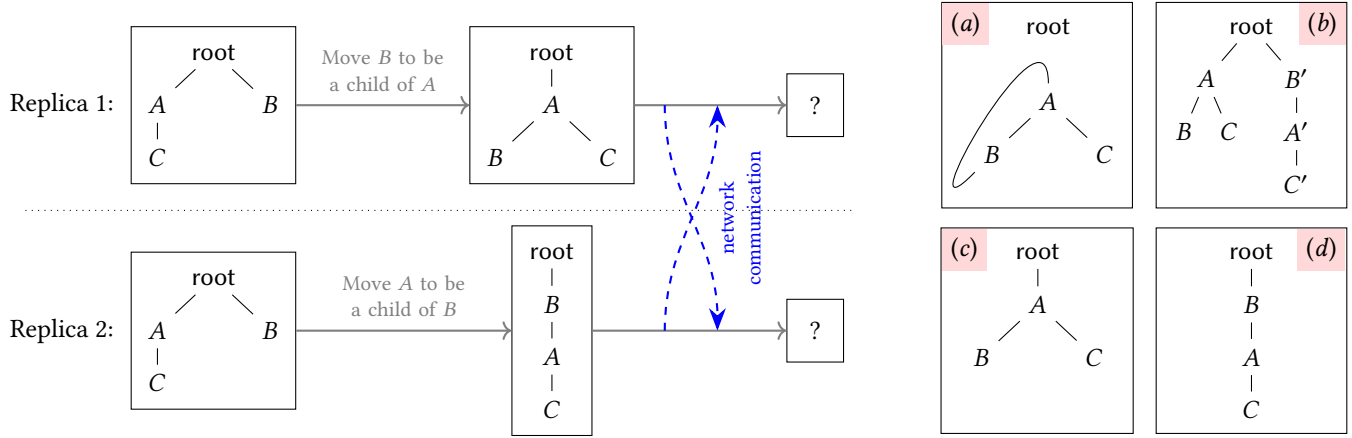
\begin{figure*}
    \centering
    \begin{tikzpicture}
        \tikzstyle{time}=[thick,->,gray]
        \tikzstyle{network}=[thick,dashed,blue,-{Stealth[length=3mm]}]
        \node [anchor=east] at (-1.2,3) {Replica 1:};
        \node [anchor=east] at (-1.2,0) {Replica 2:};
        \node [rectangle,draw] (start1) at (0,3) {
            \begin{tikzpicture}[level distance=7mm]
            \node {$\mathsf{root}$} child {node {$A$} child {node {$C$}}} child {node {$B$}};
            \end{tikzpicture}
        };
        \node [rectangle,draw] (start2) at (0,0) {
            \begin{tikzpicture}[level distance=7mm]
            \node {$\mathsf{root}$} child {node {$A$} child {node {$C$}}} child {node {$B$}};
            \end{tikzpicture}
        };
        \node [rectangle,draw] (change1) at (4.5,3) {
            \begin{tikzpicture}[level distance=7mm]
            \node {$\mathsf{root}$} child {node {$A$} child {node {$B$}} child {node {$C$}}};
            \end{tikzpicture}
        };
        \node [rectangle,draw] (change2) at (4.5,0) {
            \begin{tikzpicture}[level distance=7mm]
            \node {$\mathsf{root}$} child {node {$B$} child {node {$A$} child {node {$C$}}}};
            \end{tikzpicture}
        };
        \node [rectangle,draw,inner sep=3mm] (merge1) at (8,3) {?};
        \node [rectangle,draw,inner sep=3mm] (merge2) at (8,0) {?};
        \draw [time] (start1.east) -- node [above,text width=2.5cm,text centered,inner ysep=5pt,font=\footnotesize] {Move $B$ to be a child of $A$} (change1.west);
        \draw [time] (change1.east) -- (merge1.west);
        \draw [time] (start2.east) -- node [above,text width=2.5cm,text centered,inner ysep=5pt,font=\footnotesize] {Move $A$ to be a child of $B$} (change2.west);
        \draw [time] (change2.east) -- (merge2.west);
        \draw [network] (6.0,0) to [out=90,in=270] (6.7,3);
        \draw [network] (6.0,3) to [out=270,in=90] (6.7,0);
        \node [rotate=90,blue,font=\footnotesize,align=center] at (7.0,1.5) {network\\communication};
        \path [draw,dotted] (-2.5,1.6) -- (8.5,1.6);
        %%%%%
        \node [rectangle, draw] at (10.95,3) {
            \begin{tikzpicture}[level distance=7mm]
            \useasboundingbox (-1.3,-1.3) rectangle (1,1.2);
            \node [fill=red!15] at (-1.10,1.03) {(\emph{a})};
            \node at (0,1) {$\mathsf{root}$};
            \node (a1) {$A$} child {node (b1) {$B$}} child {node {$C$}};
            \draw (b1.south west) .. controls (-2,-2) and (0,1.5) .. (a1.north);
            \end{tikzpicture}
        };
        \node [rectangle, draw] at (13.8,3) {
            \begin{tikzpicture}[level distance=7mm]
            \useasboundingbox (-1.15,-2.3) rectangle (1.15,0.2);
            \node [fill=red!15] at (0.94,0.03) {(\emph{b})};
            \tikzstyle{level 1}=[sibling distance=12mm]
            \tikzstyle{level 2}=[sibling distance=6mm]
            \node {$\mathsf{root}$}
                child {node {$A$} child {node {$B$}} child {node {$C$}}}
                child {node {$B'$} child {node {$A'$} child {node {$C'$}}}};
            \end{tikzpicture}
        };
        \node [rectangle, draw] (start) at (10.95,0) {
            \begin{tikzpicture}[level distance=7mm]
            \useasboundingbox (-1.17,-2.3) rectangle (1.15,0.2);
            \node [fill=red!15] at (-0.98,0.03) {(\emph{c})};
            \node {$\mathsf{root}$} child {node {$A$} child {node {$B$}} child {node {$C$}}};
            \end{tikzpicture}
        };
        \node [rectangle, draw] (right) at (13.8,0) {
            \begin{tikzpicture}[level distance=7mm]
            \useasboundingbox (-1.15,-2.3) rectangle (1.15,0.2);
            \node [fill=red!15] at (0.93,0.03) {(\emph{d})};
            \node {$\mathsf{root}$} child {node {$B$} child {node {$A$} child {node {$C$}}}};
            \end{tikzpicture}
        };
    \end{tikzpicture}
    \caption{Initially, nodes $A$ and $B$ are siblings. Replica 1 moves $B$ to be a child of $A$, while concurrently replica 2 moves $A$ to be a child of $B$. Boxes (\emph{a}) to (\emph{d}) show possible outcomes after the replicas have communicated and merged their states: (a) A and B form a cycle; (b) concurrently moved subtrees are duplicated; (c) Replica 2's move is ignored; (d) Replica 1's move is ignored. Figure from~\cite{kleppmann2021highly}.}
    \label{fig:move-cycle}
\end{figure*}

Although move operations are straightforward for an unreplicated JSON object, as they only require deletion and reinsertion, they become challenging in the context of CRDTs. If several replicas concurrently delete and reinsert the same object, the merged result contains duplicates of the moved object~\cite{kleppmann2020moving}. Another issue arises with cycles, as illustrated in Figure~\ref{fig:move-cycle}. In this example, each device individually executes a move operation without creating a cycle. However, when merging these operations, a cycle may appear if the algorithm does not take care to prevent the cycle. Currently, Automerge handles moves by deletion and reinsertion, which results in behaviour (b) in Figure~\ref{fig:move-cycle}.

Previous research has demonstrated the possibility of move operations on lists and trees~\cite{kleppmann2021highly, nair2021coordination, kleppmann2020moving, najafzadeh2017co}, but the existing algorithms treat the children of a tree node as an unordered set, rendering them unsuitable for direct application to JSON trees. In JSON trees, a branch node is either a key-value map or an ordered list. Therefore, the CRDT algorithm must handle operations on those maps and lists that may interact with concurrent move operations (for example, by overwriting a key containing a node being moved). Moreover, the algorithm needs to be able to move elements between a list and a map. In this paper, we show how to implement a move operation with this functionality. We also developed novel optimisations to improve its performance. We implemented the algorithm as a standalone prototype to facilitate experimentation with various algorithm variants, showcasing the impact of these optimisations and the practical feasibility.

\begin{figure*}
    \centering
    \begin{tikzpicture}
        \node (document) [draw, text width=10em] at (0,0) {
            \begin{lstlisting}
  *@\colorbox{opset1!10}{A: a,}@*
  *@\colorbox{opset2!10}{B: [}@**@\colorbox{opset4!10}{\st{b1,}}@**@\colorbox{opset2!10}{ b2, b3],}@*
  *@\colorbox{opset3!10}{C: \{}@*
  *@\colorbox{opset3!10}{\quad D: d}@*
  *@\colorbox{opset3!10}{\}}@*
            \end{lstlisting}};

        \node (opset) [draw=none,] at (10, 0) {
            \begin{tabular}{|c|c|c|c|c|c|}
                \hline
                ID                          & Type & Object ID                  & Key                        & Value & Predecessors \\
                \hline
                \rowcolor{opset1!10} $\langle 1, 0\rangle$   & put  & $\langle 0, 0\rangle$ & A                          & a     &              \\
                \rowcolor{opset2!10} $\langle 2, 0\rangle$  & make & $\langle 0, 0\rangle$ & B                          & list  &              \\
                \rowcolor{opset3!10} $\langle 6, 0\rangle$ & make & $\langle 0, 0\rangle$ & C                          & map   &              \\
                \rowcolor{opset3!10} $\langle 7, 0\rangle$ & put  & $\langle 6, 0\rangle$ & D                          & d     &              \\
                \rowcolor{opset2!10} $\langle 3, 0\rangle$  & put  & $\langle 2, 0\rangle$ & $\langle 0, 0\rangle$ & b1    &              \\
                \rowcolor{opset2!10} $\langle 4, 0\rangle$  & put  & $\langle 2, 0\rangle$ & $\langle 3, 0\rangle$ & b2    &              \\
                \rowcolor{opset2!10} $\langle 5, 0\rangle$  & put  & $\langle 2, 0\rangle$ & $\langle 4, 0\rangle$ & b3    &              \\
                \rowcolor{opset4!10} $\langle 8, 0\rangle$  & delete &  &  &    & $\langle 3, 0\rangle$\\
                \hline
            \end{tabular}
        };

        \draw[thick,->,gray] (document) -- node[above] {Internal OpSet} (opset);

    \end{tikzpicture}
    \caption{An example JSON document with its internal OpSet}
    \label{fig:opset-example}
\end{figure*}
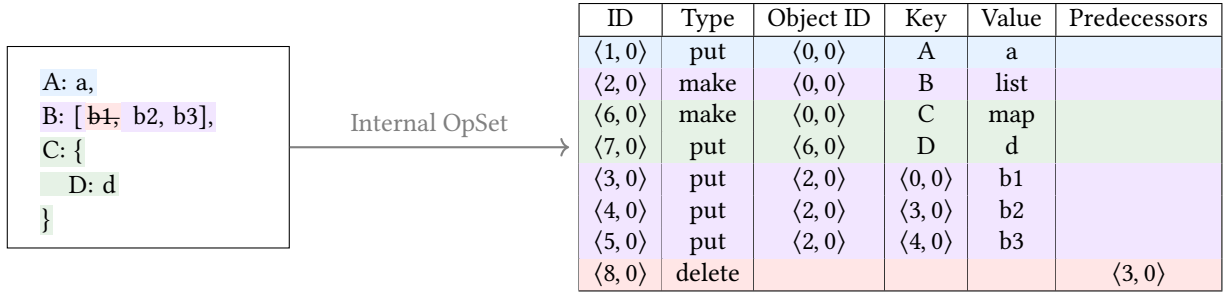

%-------------------------------------------------------------------------------
\section{The Core Mechanics of Automerge}
%-------------------------------------------------------------------------------

Automerge uses a monotonically growing set of \emph{operations} called \emph{OpSet} as its internal representation. The set of operations can be exchanged between different peers, and two sets can be merged by taking their union. Figure \ref{fig:opset-example} shows an example JSON document and its internal OpSet. 

When modifying a document, a new operation is added to the OpSet. Operations are never removed from the OpSet. An operation sets or deletes the value on a single key, or a single element in a list. 

Every operation in Automerge is assigned a unique ID, its \emph{opID}, which is implemented as a Lamport Clock~\cite{lamport2019time}. Every operation references the ID of the list or map object that it modifies, which is stored as the \emph{object ID} of the operation. A list or map is created using a ``make'' operation, and the ID of this operation subsequently serves as the unique identifier for the object it created.

If an operation $\mathit{op}_2$ overwrites, deletes or moves an existing key in a map or element in a list, and that existing value was assigned by a prior operation $\mathit{op}_1$, we say that $\mathit{op}_1$ is a \emph{predecessor} of $\mathit{op}_2$, and $\mathit{op}_2$ is a \emph{successor} of $\mathit{op}_1$. Every operation includes the opIDs of its immediate predecessors. Multiple predecessors could exist, as several prior operations may concurrently assign values to the same key. An operation is invisible if it has one or more successors.

\section{The Moving Algorithm}

When the user moves an element from one location to another, a move operation is generated and added to the OpSet. This move operation identifies the element being moved and where it is moved to. The destination of the move is determined by an object ID and a key if the destination object is a map, or a list element ID if it is a list. The ID of a prior put or make operation identifies the element being moved, and a move operation has a \emph{MoveID} field where this ID is stored. 

\subsection{Validity of Move Operations}

\begin{algorithm*}
    \caption{A naive approach for updating validity of operations}
    \label{algo:simple-validity}
    \begin{multicols}{2}
    \begin{algorithmic}[1]
    \Require ops - operations in the OpSet in ascending ID order

    \Ensure valid - a map from operation ID to its validity

    \Procedure{UpdateValidity}{ops}
    \State tree $\gets \{\}$ \Comment{a map from child ID to parent ID}
    \State winners $\gets \{\}$ \Comment{a map from object ID to winner ID}
    \State valid $\gets \{\}$
    \For{op in ops} \Comment{in order of ascending op.ID}
        \For{pred in op.Predecessors} \label{line:pred-start}
            \State tp $\gets$ pred.Type
            \If {tp == move \hspace{-0.02cm}\textbf{and}\hspace{-0.02cm} valid[pred.ID]}
                \State tree[pred.MoveID] $\gets$ null
            \ElsIf {tp == make}
                \State tree[pred.ID] $\gets$ null
            \EndIf
        \EndFor \label{line:pred-end}
        \If{op.Type == make}
            \State tree[op.ID] $\gets$ op.ObjectID
        \ElsIf{op.Type == move} \label{line:validity-start}
            \State mid $\gets$ op.MoveID
            \State oid $\gets$ op.ObjectID
            \If {\Call{IsAncestor}{tree, oid, mid}}
                \State valid[op.ID] $\gets$ false
                \State \textbf{continue}
            \EndIf
            \State valid[op.ID] $\gets$ true
            \State tree[mid] $\gets$ oid
            \State prevWinner $\gets$ winners[mid]
            \If {prevWinner != null}
                \State valid[prevWinner] $\gets$ false
            \EndIf
            \State winners[mid] $\gets$ op.ID
        \EndIf \label{line:validity-end}
    \EndFor
    \State \Return{valid} 
    \EndProcedure
    % new line
    \\
    \Function{IsAncestor}{tree, node, ancestor}
    \While{true}
        \If{node == ancestor}
        \State \Return{true}
        \EndIf
        \If{node == null || node == root}
        \State \Return{false}
        \EndIf
        \State node $\gets$ tree[node]
    \EndWhile
    \EndFunction
    \end{algorithmic}
    \end{multicols}
\end{algorithm*}

If multiple concurrent move operations move the same element, we define the move operation with the largest ID as the winner, and others have no effect. This prevents the concurrent moves from duplicating the element. A move operation is defined to be valid if and only if there is no concurrent move operation with a greater ID that moves the same element, and it does not introduce any cycles. If a move operation is invalid, the move operation itself is invisible and its predecessors remain visible (unless they have another successor operation that is valid).

The challenge is to ensure a consistent decision among replicas regarding the validity of the same operation. Consider Figure~\ref{fig:move-cycle}, where Replica 1 moves B to be a child of A and then receives the change from Replica 2, while Replica 2 moves A to be a child of B and then receives the change from Replica 1. If we check the validity at the time of applying operations, Replica 1 would consider the operation of moving A to be a child of B as invalid, while Replica 2 would say the operation of moving B to be a child of A is invalid, resulting in inconsistent document states on two replicas. 

To ensure consistent decisions on the validity of operations, we apply operations in ascending ID order. We first show a simple but inefficient approach in Algorithm~\ref{algo:simple-validity}: whenever an operation is inserted into OpSet, all the operations in the OpSet are reapplied, and the validity of operations is updated accordingly. We present an optimized algorithm in Section~\ref{sec:RAR}.

We maintain a map named \emph{tree} to keep track of the parent-child relationship between objects. All objects within a list are considered children of the list object, and the same applies to map objects. When an object is deleted from the tree, we record this in the map by setting the deleted object's parent to \emph{null}. A deletion can be thought of as similar to moving the deleted object to a ``trash'' tree that is separate from the visible document tree. 

Additionally, we maintain another map named \emph{winners}, whose key is the ID of an element being moved, and the associated value is the greatest ID among operations that move this element. 

By reapplying the operations, we update the tree to reflect any changes in the parent-child relationship in the document. Lines~\ref{line:pred-start} to~\ref{line:pred-end} set the parent to null for any objects that are deleted or overwritten. Lines~\ref{line:validity-start} to~\ref{line:validity-end} update the validity by checking for cycles and concurrent moves.

In Figure \ref{fig:move-cycle}, the result (d) implies replica 2's operation precedes replica 1's. From replica 1's perspective, its operation starts as visible but becomes invisible after receiving the operation from replica 2. It's possible to construct more complex scenarios in which an operation goes from invisible to visible as a result of receiving a remote operation.

\begin{algorithm*}
    \caption{The Restore-Apply-Reapply approach for updating validity of operations}
    \label{algo:optimized-validity}
    \begin{algorithmic}[1]
        \Procedure{UpdateValidity}{}
            \State ops $\gets []$ \Comment{operations in ascending ID order}
            \State tree $\gets \{\}$ \Comment{a map from child ID to its parent ID}
            \State moves $\gets \{\}$ \Comment{a map from element ID to a stack of move operation IDs for moving this element}
            \State parents $\gets []$ \Comment{a list, where each element is a map, from object ID to its parent ID}
            \State valid $\gets \{\}$
            \While{whenever the local replica receives an operation o}
                \State Insert o into ops at index i such that all operations at indexes > i have an ID greater than o.ID, and all operations at indexes < i have an ID less than o.ID
                \State Insert a new map into parents at index i
                \For{k $\gets$ |ops| - 1 to i + 1} \label{line:restore-start} \Comment{Restore}
                    \For {(object, location) in parents[k]}
                        \State tree[object] = location
                    \EndFor
                    \If{ops[k].Type is move \textbf{and} valid[ops[k].ID]} 
                        \State moves[ops[k].MoveID].pop()
                        \State prevMove $\gets$ moves[ops[k].MoveID].peek()
                        \If{prevMove != null}:
                            \State valid[prevMove] $\gets$ true
                        \EndIf
                    \EndIf
                \EndFor \label{line:restore-end}

                \For{k $\gets$ i to |ops| - 1} \label{line:apply-start} \Comment{Apply and Reapply}
                    \State op $\gets$ ops[k]
                    \State parent $\gets$ \{\}
                    \State prevParent $\gets$ null
                    \If{op.Type is make}
                        \State tree[op.ID] $\gets$ op.ObjectID
                        \State parent[op.ID] $\gets$ null \Comment{restore a make operation is to delete the object it creates}
                    \ElsIf{op.Type is move}
                        \If {op.Value == null} \Comment{the value is not null if it moves a scalar value} \label{line:check-value}
                            \If{\Call{IsAncestor}{tree, op.ObjectID, op.MoveID}}
                                \State valid[op.ID] $\gets$ false
                                \State \textbf{continue}
                            \EndIf
                            \State prevParent = tree[op.MoveID]
                            \State tree[op.MoveID] $\gets$ op.ObjectID
                        \EndIf
                        \State valid[op.ID] $\gets$ true
                        \If {moves[op.MoveID] == null}
                            \State moves[op.MoveID] $\gets$ new stack
                        \EndIf
                        \State prevMove $\gets$ moves[op.MoveID].peek()
                        \If{prevMove != null}:
                            \State valid[prevMove] $\gets$ false
                        \EndIf
                        \State moves[op.MoveID].push(op.ID)
                    \EndIf

                    \For{pred in op.Predecessors} \Comment{handle deleted and overwritten objects}
                        \If{pred.Type is move \textbf{and} valid[pred.ID]}
                            \State parent[pred.MoveID] $\gets$ tree[pred.MoveID]
                            \State tree[pred.MoveID] $\gets$ null
                        \ElsIf{pred.Type == make}
                            \State parent[pred.ID] $\gets$ tree[pred.ID]
                            \State tree[pred.ID] $\gets$ null
                        \EndIf
                    \EndFor
                    \If {prevParent != null}
                        \State parent[op.MoveID] $\gets$ prevParent
                    \EndIf
                    \State parents[k] $\gets$ parent

                \EndFor \label{line:apply-end}
            \EndWhile
        \EndProcedure
    \end{algorithmic}
\end{algorithm*}

\subsection{Performance Optimisation}\label{sec:RAR}

Algorithm~\ref{algo:simple-validity} ensures the consistency of move operations by applying them in the same order on all replicas. However, reapplying the entire set of operations in the OpSet is very inefficient.

To optimise the algorithm, we can avoid executing the parts that repeat the previous invocation of \emph{UpdateValidity}. Let $n$ be the ID of the newly added operation. In Algorithm~\ref{algo:simple-validity}, the execution of the loop for operations with an ID less than $n$ is the same as the previous invocation. During real-time collaboration, the operations with IDs greater than $n$ are usually a small fraction of the total set of operations.

% Besides, the make and put opeartion never generates a cycle or duplicates, which can be excluded from the process of reapplying. However, delete operation must be reapplied as they could potentially affect the cycle. Figure~\ref{fig:move-delete} shows the example where delete operation may change a move operation from invalid state to valid state. The valid move operation actually moves the target element to be a child of the deleted element, resulting the target element being deleted. 

Algorithm~\ref{algo:optimized-validity} shows an optimised version of Algorithm~\ref{algo:simple-validity}. To insert a new operation with ID $n$ into OpSet, our algorithm first restores the parent-child relationship to its state when only operations with IDs less than $n$ are applied. We then apply the new operation, followed by reapplying the operations with IDs greater than $n$. We call this sequence of steps the \emph{Restore-Apply-Reapply (RAR)} procedure.

To restore the parent-child relationship to a past state, we store former parents of overwritten, deleted, and moved objects in the variable \emph{parents}, in which we maintain a map from object IDs to the old parent IDs for each operation. 

Additionally, to restore the validity status of operations with a lower ID, we maintain a stack of move operations for every object that is moved. Operations in a stack are sorted by ID, with the top of the stack being the winner among the concurrent move operations of the same element. To restore to a past state, all the move operations with IDs greater than that of the past operation are popped.

Lines~\ref{line:restore-start} to~\ref{line:restore-end} in Algorithm~\ref{algo:optimized-validity} restore the effects of the operations greater than the new operation. Lines~\ref{line:apply-start} to~\ref{line:apply-end} update the tree to reflect any changes in the parent-child relationship by applying the newly inserted operation and reapplying the following operations. The validity is updated by detecting cycles and concurrent move operations during applying and reapplying operations similarly to Algorithm~\ref{algo:simple-validity}. Additionally, we store the old parent of objects, ensuring the ability to restore to any past state. The algorithm doesn't check for cycles if the move operation moves a scalar value, as it is impossible to create a cycle in this case.

\begin{figure}[t]
    \centering
    \includegraphics[width=1\columnwidth]{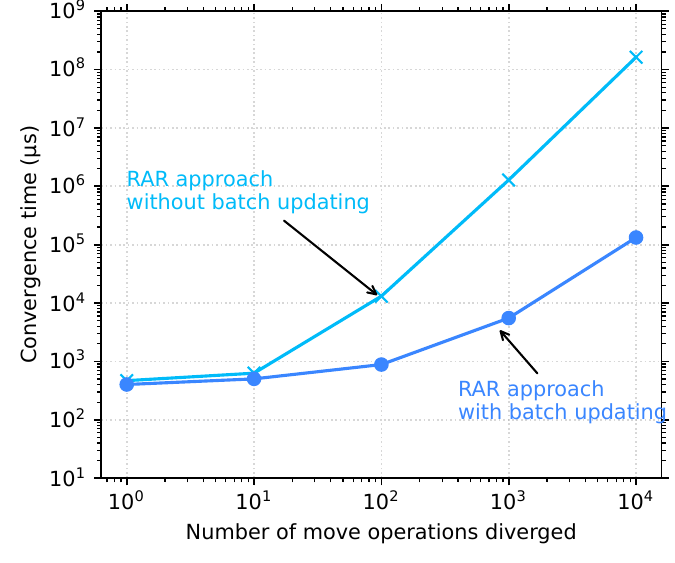}
    \caption{Convergence time of two actors that diverge by move operations}
    \label{fig:plot-convergence}
\end{figure}

\subsection{Further Optimisations}

\subsubsection{Batch Updating}\label{sec:bulk-update}

Sometimes a large number of operations are applied at once, e.g. when a user has been working offline and comes back online. Instead of calling UpdateValidity for each operation, it is more efficient to collectively apply all operations at once, and thereby amortise the cost of restoring and reapplying.

\subsubsection{Lifecycle Tracking}

In the process of Restore-Apply-Reapply, objects shift between the tree and the trash as they are moved, deleted or overwritten. As each operation's ID is a logical timestamp, we can create a sequence of IDs for each object that traces these shifts over time. We call this sequence of IDs the \emph{LifecycleList} of an object. The LifecycleList can be divided into two sublists: the PresentList, which consists of operations that create or move the object, and the TrashList, which also contains operations that overwrite or delete the object. Upon receiving an operation, we insert the ID of the new operation into the TrashList of objects it deletes or overwrites, as well as the PresentList of objects it creates or moves.

By storing the lifecycle of each object, it is no longer necessary to restore and reapply non-move operations, which merely shift objects back and forth between the tree and the trash, without changing their parents. Having to perform RAR only for move operations results in a large performance improvement for workloads where most operations create/delete objects and update values within objects, and only a small fraction of operations are moves.

\section{Evaluation}

We implemented the move algorithm in Go in a standalone prototype, which is a simplified version of Automerge. The source code is available on GitHub\footnote{\url{https://github.com/LiangrunDa/AutomergeWithMove}}. We plan to integrate the algorithm into the Rust implementation of Automerge~\cite{automerge} in the future. 
We ran the experiments on AWS c5.large instances, each having 4vCPU and 2GB RAM. 

% We deployed this on two AWS c5.large instances, with each instance having 4vCPU and 2GB RAM.

\begin{figure}[t]
    \centering
    \includegraphics[width=1\columnwidth]{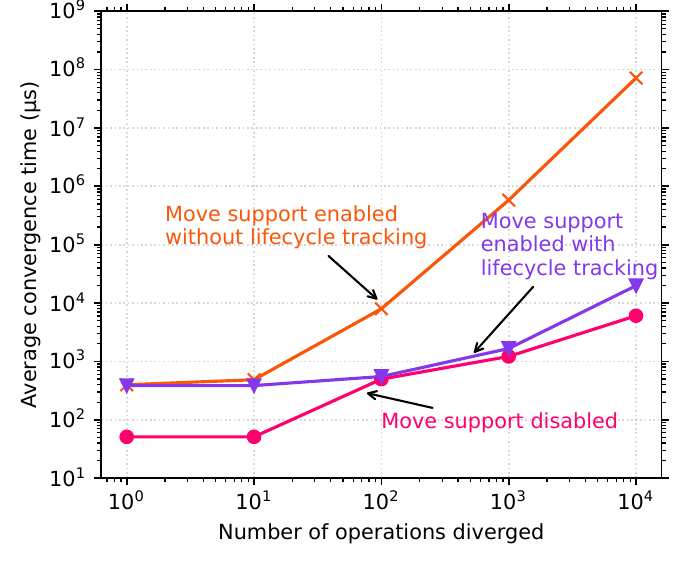}
    \caption{Convergence time of two actors that diverge by non-move operations}
    \label{fig:plot-overhead}
\end{figure}

\subsection{Convergence Complexity and Performance}

To measure the time it takes for two divergent actors to converge, two actors start with identical documents containing 100 map objects. They each generate $N$ move operations concurrently, and then send the operations to each other. For each move we choose a random object to be moved, and a random object as the destination. 

Figure~\ref{fig:plot-convergence} shows the result of the experiment. With batch updating, when N = 100, it takes about 1 ms to converge, which is acceptable for most real-time collaborative applications.

\subsection{Overhead Caused by Move Support}

Even when an application does not use any move operations, our algorithm needs to perform additional work on operations that modify the parent-child relationship of objects by creating, deleting or overwriting objects, compared to an implementation that does not support moves. To quantify this overhead, we measure the time to achieve convergence on our implementation with support for move operations enabled, and compare it to a version of the same implementation with support for move operations disabled by removing calls to UpdateValidity. Initially, both actors have identical documents containing 100 map objects. They each concurrently add a further N objects to the document, and then send the operations to each other. Batch updating is not used in this experiment.

Figure~\ref{fig:plot-overhead} shows the results. Without lifecycle tracking, the convergence time grows to more than 60 seconds for $N = 10^4$; with lifecycle tracking, this is reduced to 20ms, which is much closer to the 6ms it takes to converge with the move operation disabled. 

\subsection{Correctness Testing}

We test the correctness of the implementation by randomly generating operations on different actors to check if they converge to the same state. This approach is inspired by Jepsen~\cite{jepsen}, a framework for distributed systems verification. 

With this approach, we found several bugs during the design of the move algorithm. Those bugs were caused by not taking care of the corner cases with combinations of inputs that are rarely encountered during normal execution. Even if this approach cannot prove the correctness of the algorithm, it can uncover some subtle corner cases and provide confidence in its correctness.

\section{Related Work}

\subsection{CRDTs for Trees}

Several previous papers present designs for JSON CRDTs \cite{Kleppmann2017json,Brocco2022melda,Rinberg2022DSON}, but none of them support a move operation. On the other hand, there are several algorithms for move operations on trees, but they focus on managing the parent-child relationship, without integrating the map or list CRDT data types that occur in JSON:

\begin{itemize}
    \item Previous work by Nair et al.~\cite{nair2021coordination} proposed a conflict-free replicated tree type with move operations. They categorized these moves into up-moves (towards the root) and down-moves (away from the root); concurrent up-moves are safe while cycles caused by other move operations are resolved by ignoring some operations.

    \item Najafzadeh et al.~\cite{najafzadeh2017co} proposed a fully asynchronous file system that replaces the move operations with copy-delete operations if a cycle occurs, leading to duplication of directories. They also proposed a mostly asynchronous file system that uses locks to coordinate concurrent move operations, but this approach is not available under network partitions.

    \item Kleppmann et al.~\cite{kleppmann2021highly} proposed a move operation for trees based on an undo-do-redo algorithm, which forms the basis of the algorithm in this paper. We change the name to Restore-Apply-Reapply to avoid confusion with a user-facing undo feature.
\end{itemize}

To our knowledge, our algorithm is the first to handle the combination of a move operation with a JSON tree structure, including the overwriting keys in map objects, moving multi-value registers within map and list objects, and reordering elements within list objects. These features make the algorithm significantly more complicated. Furthermore, we introduce the concept of lifecycle tracking to reduce the overhead of the algorithm, which is not considered in previous work on move operations.

\subsection{CRDTs for Lists}

\begin{sloppypar}There are many CRDTs for lists, such as Treedoc~\cite{preguicca2009commutative}, WOOT~\cite{oster2006data}, Logoot~\cite{weiss2009logoot}, and LSEQ~\cite{nedelec2013lseq}. However, none of them support move operations.\end{sloppypar}

DSON~\cite{Rinberg2022DSON} is a delta-based JSON CRDT that supports reordering of items in lists, but not tree moves that change the parent of a node.

Kleppmann introduced an algorithm to extend existing List CRDTs with move operations~\cite{kleppmann2020moving}. The algorithm uses an LWW register for each element to track the location of the element. We incorporate the algorithm into our Restore-Apply-Reapply procedure by tracking concurrent move operations that move the same element and selecting the move operation with the greatest ID as the winner.

\section{Conclusions}

In this paper, we present an implementation of move operations in a JSON CRDT, a tree of nested map and list objects. We optimise the procedure using batch updating, and the novel technique of lifecycle tracking.

Our performance experiments demonstrate the practical feasibility of the move operation, even in scenarios with large numbers of concurrent operations. We also evaluate the correctness of our implementation through randomised testing, ensuring that the algorithm converges to the same state across different actors. In future work we plan to integrate this algorithm into the Automerge CRDT library and formally verify its correctness.

\begin{acks}
    This work was done while Martin Kleppmann was at the Technical University of Munich, funded by the Volkswagen Foundation and crowdfunding supporters including Mintter and SoftwareMill.
\end{acks}
    
\bibliographystyle{ACM-Reference-Format}

\bibliography{references}

\end{document}